\RequirePackage{fix-cm}
\documentclass[smallextended]{svjour3}        
\smartqed  
\usepackage{graphicx}
\usepackage{pdflscape}
\usepackage{amsmath}
\usepackage{amsfonts}
\newcommand{\be}{\begin{equation}}
\newcommand{\ee}{\end{equation}}

\newcommand{\ba}[1]{\left(\begin{array}{#1}}
\newcommand{\ea}{\end{array}\right)}
\journalname{Quantum Information Processing}
\usepackage{rotating}

\begin{document}
\title{Characterizing nonlocality of pure symmetric three-qubit states}


\author{K. Anjali \and Akshata Shenoy Hejamadi  \and H. S. Karthik \and S. Sahu  \and Sudha\thanks{Corresponding author: Sudha} \and A. R. Usha Devi }

\authorrunning{} 

\institute{K. Anjali \at
              Department of Physics, Bangalore University, Bangalore-560 056, India\\
  \and              
              Akshata Shenoy Hejamadi \at International Centre for Theory of Quantum Technologies, University of Gdansk, Gdansk, Poland\\  
						\and
              H. S. Karthik \at International Centre for Theory of Quantum Technologies, University of Gdansk, Gdansk, Poland \\ 
            \and 
						S. Sahu
								 \at School of Electronics and Electrical Engineering, University of Leeds, Leeds, UK
\and 
Sudha \at Department of Physics, Kuvempu University, 
	Shankaraghatta-577 451, India \\ \email{arss@rediffmail.com}\\ Inspire Institute Inc., Alexandria, Virginia, 22303, USA. \\ 
	\and
A. R. Usha Devi 
\at Department of Physics, Bangalore University, Bangalore-560 056, India\\
Inspire Institute Inc., Alexandria, Virginia, 22303, USA.}              
\date{}
\maketitle

\begin{abstract} 
We explore nonlocality of  three-qubit pure symmetric states shared between Alice, Bob and Charlie using the Clauser-Horne-Shimony-Holt (CHSH) inequality. 
We make use of the elegant parametrization in the canonical form of these states, proposed by Meill and Meyer (Phys. Rev. A {\bf 96}, 062310 (2017)) based on Majorana geometric representation.
The reduced two-qubit states, extracted from an arbitrary pure entangled symmetric three-qubit state  do not violate the CHSH inequality and hence they are  CHSH-{\emph{local}}. However, when Alice and Bob perform a CHSH test, after conditioning over measurement results of Charlie, nonlocality of the state is revealed. We have also shown that two different families of  three-qubit pure symmetric  states, consisting  of two and three distinct  spinors (qubits) respectively, can be distinguished based on the strength of violation in  the conditional CHSH nonlocality test.  Furthermore, we  identify {\em six} of the 46 classes of tight Bell inequalities in the three-party, two-setting, two-outcome i.e., (3,2,2) scenario (Phys. Rev. A 94, 062121 (2016)). Among the two inequivalent families of three-qubit pure symmetric states, {\em only} the states belonging to three distinct spinor class show maximum violations of these six tight Bell inequalities. 
\keywords{Bell-CHSH non-locality test \and Conditional CHSH test \and (3,2,2) Scenario \and Permutation symmetric three
qubit states  \and Two and three distinct spinor classes}
\PACS{03.65.Ud, 03.67.-a}

\end{abstract}
\section{Introduction} 

Correlations arising from local measurements made on a spatially separated composite quantum system are {\emph{nonlocal}} if the underlying measurement statistics cannot be reproduced using classical strategies. Violation of Bell inequalities~\cite{bell1,chsh} certifies the presence of  nonlocality. There has been a
growing demand for more general ways of characterizing nonlocality and over the years different theoretical formulations of Bell inequalities have been proposed~\cite{cavalcanti,rchaves,rmp,lewenstein1,lewenstein2} to capture nonlocality exhibited by shared multipartite quantum states. Composite quantum states which violate  Bell inequality are necessarily entangled but the converse is not necessarily true.  Characterization of entanglement and nonlocality of multipartite quantum systems has gained  growing importance as they  play a key role in several quantum information processing tasks~\cite{rmp}.     

Multiqubit states which respect exchange symmetry form an important class of composite quantum systems.  Examples of these states are the well known GHZ and W states~\cite{aditimark,bastin,aru}. Permutation symmetric multiqubit states have  attracted much attention in view of their applications in quantum information processing.  Entanglement properties of symmetric multiqubit states have been studied extensively~\cite{aditimark,bastin,markham,aru,toth,aru1,aru2,aru3}. Confining to the special class of permutation symmetric states leads to reduction in the number of parameters and this in turn helps in simplifying evaluations in an explicit manner.  Moreover, nonlocality of multiqubit states obeying exchange symmetry have been investigated using {\em experimentalist-friendly}  Bell inequalities involving multiqubit observables exhibiting  one and two qubit correlations only~\cite{lewenstein1,lewenstein2}.  
 
Recently Meill and Meyer~\cite{meyer} employed a canonical form for permutation symmetric pure entangled three-qubit states based on the Majorana geometric representation~\cite{bastin,aru,majorana} and evaluated the algebraically independent local unitary invariant quantities. Majorana geometric representation offers a natural classification of entangled pure  symmetric three-qubit states, consisting of two and  three distinct spinors (qubits), which can be represented by distinct  points  on the Bloch sphere. This structure leads to  an elegant parametrization for the entangled pure symmetric three-qubit states~\cite{meyer} and facilitates explicit evaluation of local unitary invariants viz., the pairwise concurrence $C$, the three-tangle $\tau$ and the Kempe invariant $\kappa$. It would be interesting to explore nonlocality features of the entire family of  entangled pure permutation symmetric three-qubit  states using their canonical structure -- equipped with a neat parametrization, resulting  from the Majorana geometric representation. We focus here on  investigating nonlocality of entangled pure symmetric three-qubit states using conditional CHSH inequality~\cite{cavalcanti,rchaves} and the tight Bell inequalities of the (3,2,2) scenario (i.e., Bell tests in the three-party, two-setting, two-outcome scenario)~\cite{pit01,silwa,cab1,cab2}.

We begin by verifying that the reduced two-qubit density matrices -- extracted from the entangled pure  symmetric three-qubit states -- are  CHSH-local. It is pertinent to recall here that Popescu and Rohrlich had shown~\cite{PR} that a bipartite system   prepared by carrying out projective measurements on the $N-2$ parties of any pure entangled $N$-party state is nonlocal.   
This prompts us to study violation of the  {\em conditional} CHSH inequality~\cite{cavalcanti,rchaves} by two-qubit states obtained after an outcome of a projective measurement on the third qubit is recorded. This provides a succinct illustration that  non locality of the reduced two-qubit state indeed gets activated by conditioning over the outcomes of measurement made on the third qubit, thereby offering a preview of the nonlocality content of the global pure entangled symmetric three-qubit state. We show that the optimal violation of conditional CHSH inequality is different for the classes of entangled pure symmetric three-qubit states,  consisting of two and  three distinct spinors respectively.  Furthermore, we show that  {\em six} of the  46 classes of tight Bell inequalities in the (3,2,2) scenario get  violated maximally by  pure symmetric  three-qubit states belonging to the three distinct spinor Majorana class.    

This paper is organized as follows: Section~2 gives a brief introduction of the Majorana geometric representation~\cite{bastin,aru,meyer,majorana} of pure entangled three-qubit symmetric states  and their classification under stochastic local operations and classical communications (SLOCC) into two different classes. We employ the parametrization given in Ref.~\cite{meyer} to characterize these two SLOCC inequivalent classes. In Section 3,  optimal expectation values of the Bell-CHSH operator are explicitly evaluated (in terms of the eigenvalues of the  two-qubit correlation matrices~\cite{horodecki95}) for the SLOCC inequivalent classes of pure symmetric three-qubit states. It is demonstrated  that the CHSH inequality is not violated by any pairs of qubits extracted from an entangled pure three-qubit symmetric state. In Section 4, violation of conditional CHSH inequality  by the states under consideration is illustrated.   Bell inequalities of the (3,2,2) scenario are discussed in Section 5 and a concise summary of our results is given in Section 6.

\section{Three-qubit pure symmetric states: SLOCC inequivalent classes} 
An arbitrary pure symmetric three-qubit state can be expressed in the Majorana geometric representation~\cite{bastin,aru,majorana} by      
\begin{eqnarray}
\vert\Psi_{\rm sym}\rangle&=&{\cal N} \sum_P \hat{P}\{ \vert \alpha_1, \beta_1\, \rangle \otimes  |\alpha_2, \beta_2\rangle\otimes \vert \alpha_3, \beta_3\rangle \}
\end{eqnarray} 
where, 
\be
\vert \alpha_k, \beta_k\rangle\equiv \cos\frac{\beta_k}{2}\,  
\vert 0\rangle + e^{i\alpha_k}\,\sin\frac{\beta_k}{2}\, \vert 1\rangle,\ \ k=1,\,2,\,3,
\ee
denote qubit states where $0\leq\alpha_k\leq 2\pi,$ $0\leq\beta_k\leq \pi$. Here $\hat{P}$ corresponds to the set of all 
permutations on the constituent qubits and ${\cal N}$ denotes the normalization factor. 

Depending on the number of distinct spinor states $\{ \vert \alpha_k, \beta_k\,\rangle,\, k=1,2,3\}$   there arise two different classes of entangled pure symmetric states of three qubits~\cite{aru}:    
\begin{enumerate}
	\item Two distinct spinor class:  $\vert\alpha_1,\beta_1\rangle =\vert\alpha_2,\beta_2\rangle\neq \vert\alpha_3,\beta_3\rangle$ 
	\item Three distinct spinor class:  $\vert\alpha_1,\beta_1\rangle \neq \vert\alpha_2,\beta_2\rangle \neq   
	\vert\alpha_3,\beta_3\rangle$. 
\end{enumerate} 
It may be noted that when the constituent qubit states are all identically equal i.e., $\vert\alpha_1,\beta_1\rangle =\vert\alpha_2,\beta_2\rangle= \vert\alpha_3,\beta_3\rangle=\vert\alpha,\beta\rangle$ the pure  state $\vert\Psi_{\rm sym}\rangle$ reduces to a separable (product) form $\vert\,\alpha,\beta\rangle^{\otimes\,3}$.

Let us denote the two distinct spinor class of pure three-qubit states by $\{{\cal{D}}_{3,2}\}$ and the three distinct spinor class by $\{{\cal{D}}_{3,3}\}$. These two classes  are evidently inequivalent under SLOCC~\cite{aru}. Examples of three-qubit states of the class  $\{{\cal{D}}_{3,2}\}$ are 
\begin{eqnarray}
\label{W}
\vert{\rm W}\rangle&=& {\cal N}\, \sum_P \hat{P}\{ \vert 0\,\rangle \otimes  |0\,\rangle\otimes \vert 1\,\rangle \}\nonumber \\
&=& \frac{1}{\sqrt{3}}\, \left(\vert 0\,0\, 1\rangle+\vert 0\,1\, 0\rangle+\vert 1\,0\, 0\rangle\right)
\end{eqnarray}
and  its obverse state $\vert\bar{{\rm W}}\rangle= \frac{1}{\sqrt{3}}\, \left(\vert 1\,1\, 0\rangle+\vert 1\,0\, 1\rangle+\vert 0\,1\, 1\rangle\right)$. 
A paradigmatic example of the three-qubit symmetric state belonging to the SLOCC class  $\{{\cal{D}}_{3,3}\}$ is the GHZ state~\cite{GHZ}:  
\begin{eqnarray}
\label{GHZ}
\vert{\rm GHZ}\rangle&=& {\cal N}\, \sum_P \hat{P}\{ \vert\, \phi\rangle_1 \,  \otimes \vert\, \phi\rangle_2  \otimes \vert \phi\rangle_3 \}\nonumber \\
&=& \frac{1}{\sqrt{2}}\, \left(\vert 0\,0\, 0\rangle+\vert 1\,1\, 1\rangle\right) 
\end{eqnarray}
where    $\vert\phi\rangle_p=\frac{1}{\sqrt{2}}(\vert 0\rangle+\omega^p\,\vert 1\rangle), p=1,2,3$ are the constituent qubit states and $\omega$ denotes cube root of unity. 
Another illustrated example of three-qubit state belonging to the class $\{{\cal{D}}_{3,3}\}$ is the superposition of  $\vert{\rm W}\rangle$ and $\vert\bar{{\rm W}}\rangle$ states~\cite{aditimark,aru}: 
\begin{eqnarray}
\vert{\rm W}\,\bar{{\rm W}}\rangle&=& {\cal N}\, \sum_P \hat{P}\{ \vert 0\,\rangle \otimes  \vert 1\,\rangle\otimes \vert +\rangle \} 
\end{eqnarray}
where $\vert +\rangle=\frac{1}{\sqrt{2}}(\vert 0\rangle+\vert 1\rangle)$.

It has been shown that  any arbitrary three-qubit pure state of two distinct spinor class $\{{\cal{D}}_{3,2}\}$ takes the following simple form  with the help of local unitary operations~\cite{meyer}: 
\begin{eqnarray}
\label{2meyer}
\vert\Psi_{3,\,2}\rangle&=&{\cal N} \sum_P \hat{P}\{ \vert 0\rangle \otimes  \vert 0\rangle \otimes \vert \beta\rangle\}
\end{eqnarray}	
where 
\be
\label{beta32}	
\vert \beta\rangle=\cos\,\frac{\beta}{2}\, \vert 0\rangle+\sin\,\frac{\beta}{2}\,\vert 1\rangle 
\ee
Thus the states belonging to the class $\{{\cal{D}}_{3,2}\}$ are characterized by {\em one} real parameter specified by  $0 < \beta \leq \pi$. Note that the state $\vert\Psi_{3,\,2}\rangle$ reduces to the three-qubit $W$ state when $\beta=\pi$.

An arbitrary three-qubit pure symmetric state belonging to the class $\{{\cal{D}}_{3,3}\}$  of three distinct spinor is shown~\cite{meyer} to be  characterized by three real parameters: 
\begin{eqnarray}
\label{3meyer} 
\vert\Psi_{3,3}\rangle&=&{\cal N} \left(\vert 0\,\rangle^{\otimes\,3}  + y\,  
e^{i\, \alpha} \, \vert \beta\rangle^{\otimes\,3} \right) 
\end{eqnarray} 
where the qubit state $\vert\beta\rangle$ is given by (\ref{beta32});  the real parameters $y$ and $\alpha$ are bounded by  $0<y\leq 1, \ 0\leq \alpha \leq 2\pi$.  
We make use of this explicit parametrization of three-qubit symmetric states for exploring their nonlocality.

\section{Bell-CHSH nonlocality test}  

Consider the reduced two-qubit  density matrix of  a pure symmetric three-qubit state $\vert\Psi\rangle_{\rm sym}$ shared by Alice, Bob and Charlie:   
\begin{eqnarray*}
	\rho_{\rm R}&=&{\rm Tr}_{A}\,\left(\vert \Psi\rangle_{\rm sym}\langle\Psi\vert\right)
	={\rm Tr}_{B}\,\left(\vert \Psi\rangle_{\rm sym}\langle\Psi\vert\right) \\ 
	&=& {\rm Tr}_{C}\,\left(\vert \Psi\rangle_{\rm sym}\langle\Psi\vert\right)  
\end{eqnarray*}
In the Hilbert-Schmidt basis we express
\begin{eqnarray}
\label{Tconv} 
\rho_{\rm R}&=& \frac{1}{4}\left[I\otimes I+\sum_{i=1}^{3}\,s_i\left(\sigma_i\otimes I+I \otimes \sigma_i \right) \right. \nonumber \\ 
& & \left. +\sum_{i,j=1}^{3}\,t_{ij}\, (\sigma_i \otimes \sigma_j)\right]
\end{eqnarray}
where  $\sigma_i$, $i=1,\,2,\,3$ are the Pauli  matrices; $I$ denotes $2\times 2$ identity matrix and the real parameters $s_i$, $t_{ij}$ are given by 
\begin{eqnarray} 
s_i&=& {\rm Tr}\left[\rho_R\, (\sigma_i\otimes I)\right] = {\rm Tr}\left[\rho_R\, (I\otimes \sigma_i)\right] \nonumber \\
\label{tbell}
t_{ij}&=&{\rm Tr}\left[\rho_R\, (\sigma_i\otimes \sigma_j)\right]=t_{ji}.
\end{eqnarray}
The  correlation matrix $T=(t_{ij})$ of the permutation symmetric density matrix $\rho_R$  is a $3\times 3$ real symmetric matrix satisfying the condition~\cite{aru1,aru2} 
${\rm Tr}(T)=1$.

We consider the CHSH inequality~\cite{chsh} given by 
\be
\label{bellineq}
\langle {\rm CHSH}\rangle=\vert\langle A_1\,B_1 +A_1\, B_2+ A_2\,B_1 -A_2\,B_2\rangle\vert\leq 2  
\ee
where 
\be
\langle A_i\, B_j\rangle=\sum_{a_i,b_j=\pm 1}\,a_i\,b_j\ p(a_i,b_j\vert A_i, B_j), \ \ i,\, j=1,\,2. 
\ee 
Here $p(a_i,\,b_j\vert A_i, B_j)$,  $i,\,j=1,2$, denote the correlation probabilities evaluated based on the outcomes $a_i$, $b_j$ of the observables $A_i$, $B_j$ of  Alice and Bob respectively.  

Maximum value  $\langle {\rm CHSH}\rangle_{\rm opt}$ achievable by any arbitrary two-qubit state is given by~\cite{horodecki95} 
\be 
\label{bineq}
\langle {\rm CHSH}\rangle_{\rm opt}=2\sqrt{t_1^2+t_2^2}
\ee
where $t_1^2$, $t_2^2$ denote two largest eigenvalues of $T^\dag\, T$.   

The parametrization (\ref{2meyer}), (\ref{3meyer}) of the pure three-qubit symmetric states facilitates evaluation of  the correlation matrix $T$  of the reduced two-qubit state $\rho_{R}$  belonging to the  SLOCC inequivalent classes $\{{\cal D}_{3,2}\}$, $\{{\cal D}_{3,3}\}$. Utilizing this structure one may compute the optimal CHSH value to test nonlocality. We carry out such an analysis in the following subsections.  

\subsection{Bell-CHSH test on the two-qubit states of the SLOCC class $\{{\cal D}_{3,2}\}$} 

We evaluate the correlation matrix $T$ (see  (\ref{tbell})) of the two-qubit reduced density matrix $\rho_R$ associated with the one-parameter family  (\ref{2meyer}) of  pure symmetric states belonging to the SLOCC class $\{{\cal D}_{3,\,2}\}$ explicitly: 
\be
T=\frac{1}{3(2+\cos \beta)}\ba{cccc} 1-\cos \beta & 0 & 3 \sin \beta \\
0 & 1-\cos \beta & 0 \\
3 \sin \beta  & 0 & 4+5 \cos \beta     \ea 
\ee  
and obtain its eigenvalues~\footnote{In the case of a symmetric two-qubit density matrix we have $T=T^\dag$. Thus, the  eigenvalues $ t_1^2,\, t_2^2,\, t_3^2$ of  $T^\dag\, T$ are determined by those of the real symmetric matrix $T$ itself.}  
\begin{eqnarray}
\label{t123} 
t_1 &=&\frac{5+4\cos \beta+3\sqrt{5+4\cos \beta}}{6(2+\cos\beta)}  \nonumber\\
t_2&=&\frac{1-\cos \beta}{3(2+\cos \beta)} \\
t_3&=&\frac{5+4\cos \beta-3\sqrt{5+4\cos \beta}}{6(2+\cos\beta)}.\nonumber
\end{eqnarray} 
We have plotted the absolute values of $t_1, t_2$ and $t_3$ in Figure~1 from which it is readily seen  that $\vert t_1\vert \geq \vert t_2\vert \geq \vert t_3\vert$.   
\begin{figure}[h]
	\label{1}
	\begin{center}
		\includegraphics*[width=3in,keepaspectratio]{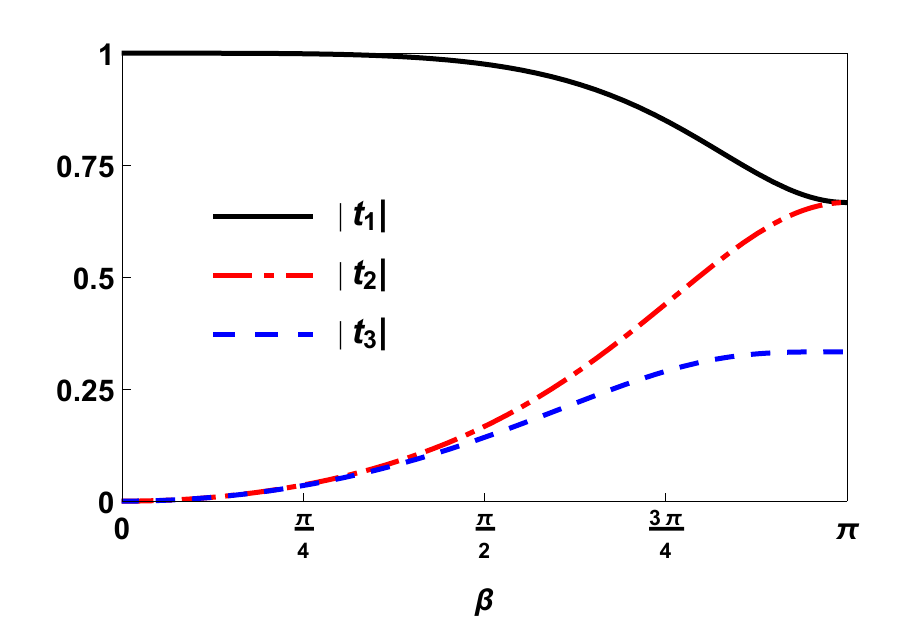}
		\caption{(Color online) Absolute values of $ t_1$ (black,solid), $t_2$ (red, dot-dashed) and  $t_3$ (blue, dashed) (see (\ref{t123})) as a function of the parameter $\beta$}.
	\end{center}
\end{figure}   
\begin{figure}[h]
	\label{2}
	\begin{center}
		\includegraphics*[width=3in,keepaspectratio]{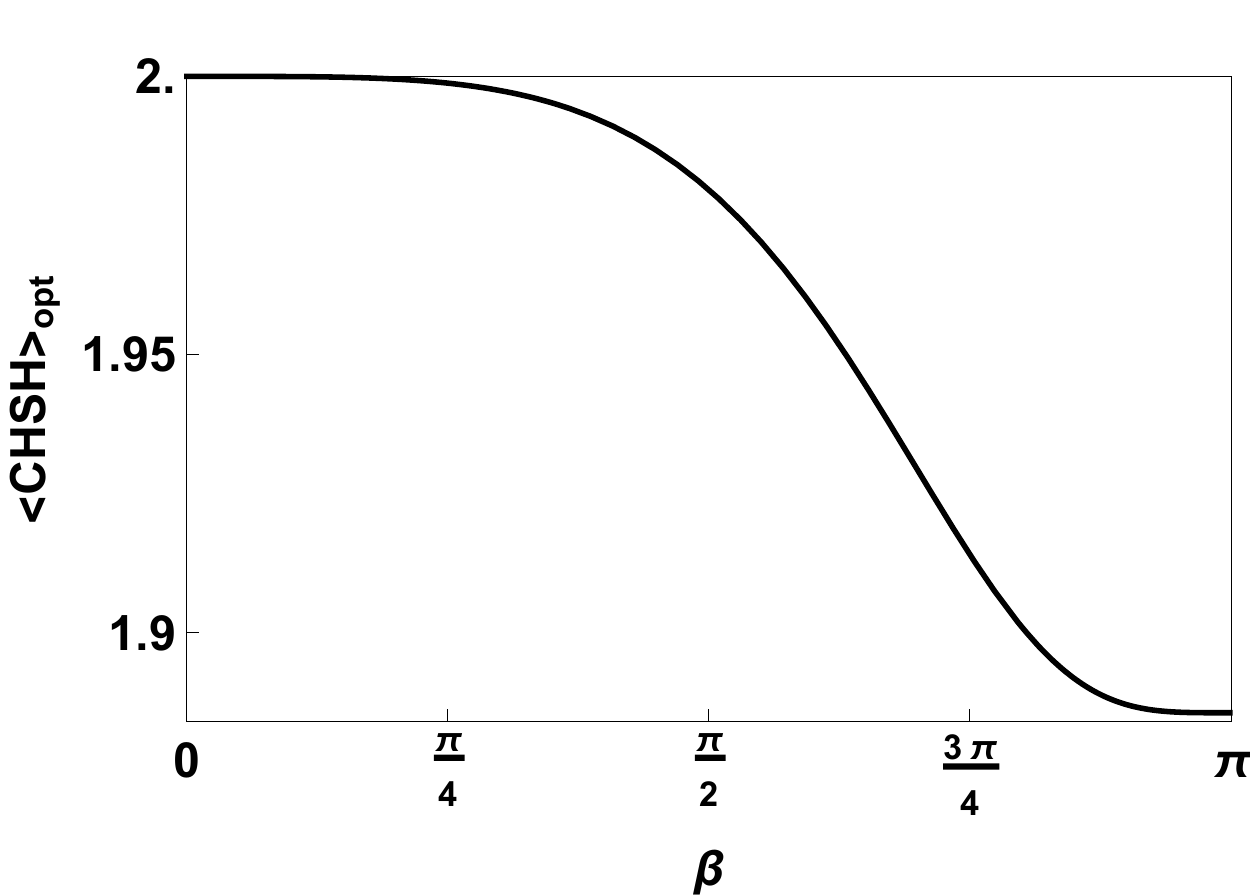}
		\caption{Optimal value $\langle{\rm CHSH}\rangle_{\rm opt}$ achievable in the two-qubit reduced state, associated with the class     
			$\{{\cal{D}}_{3,2}\}$ of entangled pure symmetric three-qubit states,  as a function of  $\beta$.} 
	\end{center}
\end{figure} 
Figure~2 depicts a plot of  $\langle {\rm CHSH}\rangle_{\rm opt}$ (See (\ref{bineq})) as a function of  $\beta$. Evidently, $\langle {\rm CHSH}\rangle_{\rm opt}\leq 2$ in the entire range of the parameter  $\beta$. This leads to an explicit verification of the result  that reduced two-qubit states of entangled pure symmetric three-qubit states belonging to the class $\{{\cal{D}}_{3,2}\}$ are CHSH-local. 

\subsection{Bell-CHSH test on the two qubit states of  $\{{\cal D}_{3,3}\}$} 
We now proceed to study the Bell-CHSH nonlocality exhibited by a pair of qubits extracted from  the three-qubit states belonging to the  SLOCC class $\{{\cal D}_{3,3}\}$.  Towards this end, we proceed to evaluate the  eigenvalues $\vert t_1\vert\geq \vert t_2\vert \geq \vert t_3\vert$  of  the  correlation matrix $T$ (see (\ref{tbell})) so as to compute the optimal CHSH value $\langle{\rm CHSH}\rangle_{\rm opt}=2\sqrt{t_1^2+t_2^2}$. 
  
The state $\vert\Psi_{3,\,3}\rangle\in\{{\cal D}_{3,3}\}$ (see (\ref{3meyer})) is characterized by three real independent parameters: $0<\beta \leq \pi$,  $0<y\leq 1$, $0\leq \alpha \leq 2\pi$.  We have verified numerically that the two-qubit states $\rho_R$ extracted from $\vert\Psi_{3,\,3}\rangle\in\{{\cal D}_{3,3}\}$ are all local. We choose two  specific sets of parameters $(y, \alpha,\beta)$ for the purpose of explicit demonstration. 

Listed below are two specific choices of parameters $(\alpha,\beta)$ and two largest eigen values $t_1,t_2$ of the correlation matrix $T$ of the associated reduced two-qubit states:  
\begin{itemize} 
	\item[(i)]  $\beta=\pi/2$, $\alpha=\pi$ 
	\begin{eqnarray}
	t_1&=&\frac{1-\sqrt{2}y+y^2+\sqrt{1+y^4}}{2-\sqrt{2}y+2y^2}, \nonumber \\  
	t_2&=&\frac{y}{\sqrt{2}(1+y^2)-y}
	\end{eqnarray}
	\item[(ii)] $\beta=\pi/3$, $\alpha=0$ 
	\be
	t_{1,\,2}=\frac{2(1+\sqrt{3}y+y^2\pm\sqrt{1+\sqrt{3}y+2y^2+\sqrt{3}y^3+y^4})}{4+3\sqrt{3}y+4y^2}
	\ee
\end{itemize}  
We have plotted  $\langle {\rm CHSH}\rangle_{\rm opt}=2\, \sqrt{t_1^2+t_2^2}$ as a function of the state parameter $y$ in the cases (i) $\beta=\pi/2$, $\alpha=\pi$ and (ii) $\beta=\pi/3$, $\alpha=0$  in Figure~3. 
\begin{figure}[h]
	\label{3}
	\begin{center}
		\includegraphics*[width=3in,keepaspectratio]{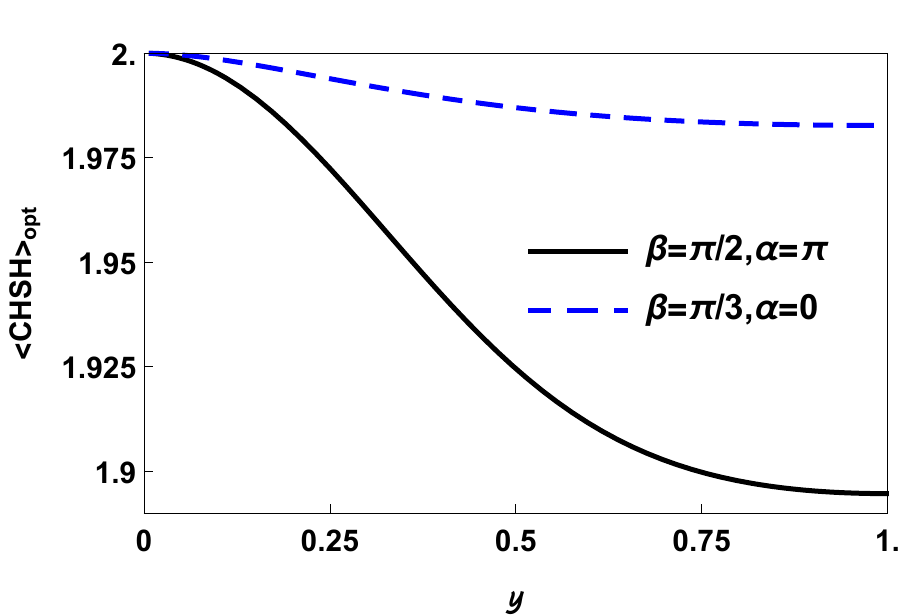}
		\caption{(Color online) Optimal CHSH value  $\langle {\rm CHSH}\rangle_{\rm opt}$ that can be realized in the two-qubit reduced state of $\vert\Psi_{3,3}\rangle$ (see (\ref{3meyer}))  belonging to the three distinct spinor class $\{{\cal{D}}_{3,3}\}$ as a function of the parameter $y$  for specific cases of parameters (i) $\beta=\pi/2,\,\alpha=\pi$ (black, solid) and (ii) $\beta=\pi/3,\,\alpha=0$ (blue, dashed) } 
	\end{center}
\end{figure}
This demonstrates that the maximum  CHSH value  $\langle {\rm CHSH}\rangle_{\rm opt}$ evaluated in a reduced two-qubit state drawn from three-qubit states $\vert \Psi_{3,3}\rangle$ (see (\ref{3meyer})) of the three distinct spinor class $\{{\cal{D}}_{3,3}\}$ is less than or equal to 2 and thus the state is local. 

\section{Conditional CHSH  test} 
In Ref.~\cite{PR}  Popescu and Rohrlich considered a generalized CHSH inequaltiy involving correlations between two systems, conditioned on a particular measurement outcome registered by the $N-2$ systems. They  proved  that any  bipartite system  prepared by carrying out an optimal projective measurements on the $N-2$ parties of any pure entangled $N$-party state violated the conditional CHSH inequality. More recently, the CHSH violation of two out of the $N$ qubits conditioned on a measurement outcome of all other
$N-2$ qubits has been employed~\cite{cavalcanti,rchaves} to certify the nonlocality of multipartite quantum states. In this section we investigate the violation of the conditional CHSH inequality by entangled pure symmetric three-qubit states.  
                 
 Let us consider an entangled pure symmetric three-qubit state, shared by Alice, Bob and Charlie. Charlie performs a projective measurement $P_c$ on his qubit, which yields a particular outcome $c=\pm 1$ (with probability $p(c)$). Conditioned on Charlie's outcome, Alice and Bob perform  local measurements (two settings each) of the observables  $A_i,\, B_j,\, i,j=1,2$  on their respective qubits and get dichotomic  outcomes $a_i, b_j=\pm 1$.  The conditional CHSH inequality is given by~\cite{cavalcanti,rchaves} 
 \be 
\label{chsh_c}
\left\vert \langle\,A_1\,B_1\rangle_c  +\langle\,A_1\,B_2\rangle_c+\langle\,A_2\,B_1\rangle_c-\langle\,A_2\,B_2\rangle_c\right\vert \leq 2\, p(c)
\ee
where
\be
\langle A_i\, B_j\rangle_c=p(c)\, \sum_{a_i,b_j=\pm 1}\,a_i\,b_j\,p(a_i,b_j\vert A_i, B_j, c).
\ee
Summation over $c=\pm 1$ leads to the following  inequality (see Eq.~(14) of Ref.\cite{rchaves}):
\be
\label{cdchsh1}
\sum_{c=\pm 1}\,  \left\vert \langle\,A_1\,B_1\rangle_c  +\langle\,A_1\,B_2\rangle_c+\langle\,A_2\,B_1\rangle_c-\langle\,A_2\,B_2\rangle_c \right\vert \leq 2\,  
\ee
In a given pure entangled three qubit symmetric state $\vert\Psi_{\rm sym}\rangle$, Charlie chooses a suitable local measurement $P_c$ and performs a projective measurement on his qubit to obtain a particular outcome $c=\pm 1$. Charlie's measurement,  in turn,  projects the two-qubit state of Alice \& Bob  such that the conditional CHSH inequality given by (\ref{cdchsh1}) gets violated. Such violation clearly demonstrates activation of hidden-nonlocality~\cite{gisin}.

In order to verify optimal violation of the conditional CHSH inequalities (\ref{chsh_c}) and (\ref{cdchsh1})  we need to compute  two  largest eigenvalues $t_1^c, t_2^c$  of the two-qubit conditional correlation matrix $T^c,\, c=\pm 1,$ so that    
\begin{eqnarray}
\label{cdc}
\langle {\rm CHSH}\rangle_{c,\rm opt}&=&2\, \sqrt{(t^c_1)^2+(t^c_2)^2} > 2 \\ 
\label{cdopt}
\langle {\rm CHSH}\rangle^{\rm con}_{\rm opt}&=& 2\, \sum_{c=\pm 1}\, p(c)\, \langle {\rm CHSH}\rangle_{c,\rm opt} \nonumber \\ 
&=&  2\, \sum_{c=\pm 1}\, p(c)\, \sqrt{(t^c_1)^2+(t^c_2)^2} > 2  
\end{eqnarray}
where 
\be
\label{pc}
p(c)=\langle\Psi_{\rm sym}\vert \,(I \otimes I \otimes P_c)\,\vert \Psi_{\rm sym}\rangle    
\ee
denotes  probability of obtaining the outcome  $c=\pm 1$ by Charlie for his local projective measurement $P_c$.  

Note that the matrix elements $t^{c}_{ij}$ of the two-qubit correlation matrix $T^c$ are given by 
\be
\label{tc}
t^c_{ij}=\langle\Psi_{\rm sym}\vert \sigma_i \otimes \sigma_j \otimes P_c\,\vert \Psi_{\rm sym}\rangle, \ \  i,j=1,2,3. 
\ee
We denote Charlie's projective measurement by  
\begin{eqnarray}
\label{pcop}
P_c(\theta, \phi)&=&\frac{1}{2}\, \left(I +c\ \vec{\sigma}\cdot \hat{n}(\theta, \phi)\,\right),    
\end{eqnarray} 
where $\hat{n}(\theta, \phi)~=~(\sin\,\theta \cos\,\phi,\, \sin\,\theta \sin\,\phi, \, \cos\, \theta)$, $0~\leq~\theta~\leq~\pi,\ \ 0~\leq~\phi~\leq~2\pi$  denotes a unit vector corresponding to the measurement setting of Charlie.  

\subsection{Violation of the conditional CHSH inequality by states of the SLOCC class $\{{\cal{D}}_{3,2}\}$}

We proceed to compute optimal conditional CHSH values $\langle {\rm CHSH}\rangle_{c,{\rm opt}},\, \langle {\rm CHSH}\rangle^{\rm con}_{\rm opt}$ (see (\ref{cdc}),(\ref{cdopt})) in the one parameter family of states $\vert \Psi_{3,2}\rangle$ given by (\ref{2meyer}) which belong to the two distinct spinor SLOCC class $\{{\cal{D}}_{3,2}\}$.

On explicit evaluation, elements of the conditional correlation matrix $T^{c}$  (See (\ref{tc})) corresponding to states $\vert \Psi_{3,2}\rangle$ are obtained as
\begin{eqnarray}
t^c_{11}&=& \left[(1+c\, \cos\theta)(1-\cos\beta)\right]/\mu_c 
=t^c_{22}\   \nonumber \\ 
t^c_{12}&=&t^c_{21}=0 \nonumber \\
t^c_{13} &=& \left[3(1+c \cos\theta)\sin\beta + c\,\cos\phi\sin\theta(1-\cos\beta)\right]/\mu_c=t^c_{31} \nonumber \\ 
t^c_{23} &=& \left[ c\, \sin\theta\,\sin\phi\,(1-\cos\beta)\right]/\mu_c=t^c_{32},\ \   \nonumber \\ 
t^c_{33}&=&1-2\, t^c_{11} 
\end{eqnarray}
where we have denoted
\begin{eqnarray}
\mu_c&=&6+5\, c\, \cos\theta+(3+4\,c\, \cos\theta)\, \cos\beta+3\,c\,\cos\phi\,\sin\beta\,\sin\theta. 
\end{eqnarray}

Substituting (\ref{pcop}) and (\ref{2meyer}) in (\ref{pc}) we obtain  Charlie's probability of getting the measurement outcome $c=\pm 1$ in the state  
$\vert \Psi_{3,2}\rangle$: 
\begin{equation}
p(c)=\frac{\mu_c}{6(2+\cos\beta)}
\end{equation}
Eigenvalues of the correlation matrix $T^c$ are given by 
\begin{eqnarray}
t^c_{1}= 1,\   t^c_2=\frac{(1+c\cos\theta)(1-\,\cos\beta)}{\mu_c} =-t^c_3  
\end{eqnarray}  

\begin{figure}[h]
		\label{4}
				\begin{center}
		\includegraphics*[width=3in,keepaspectratio]{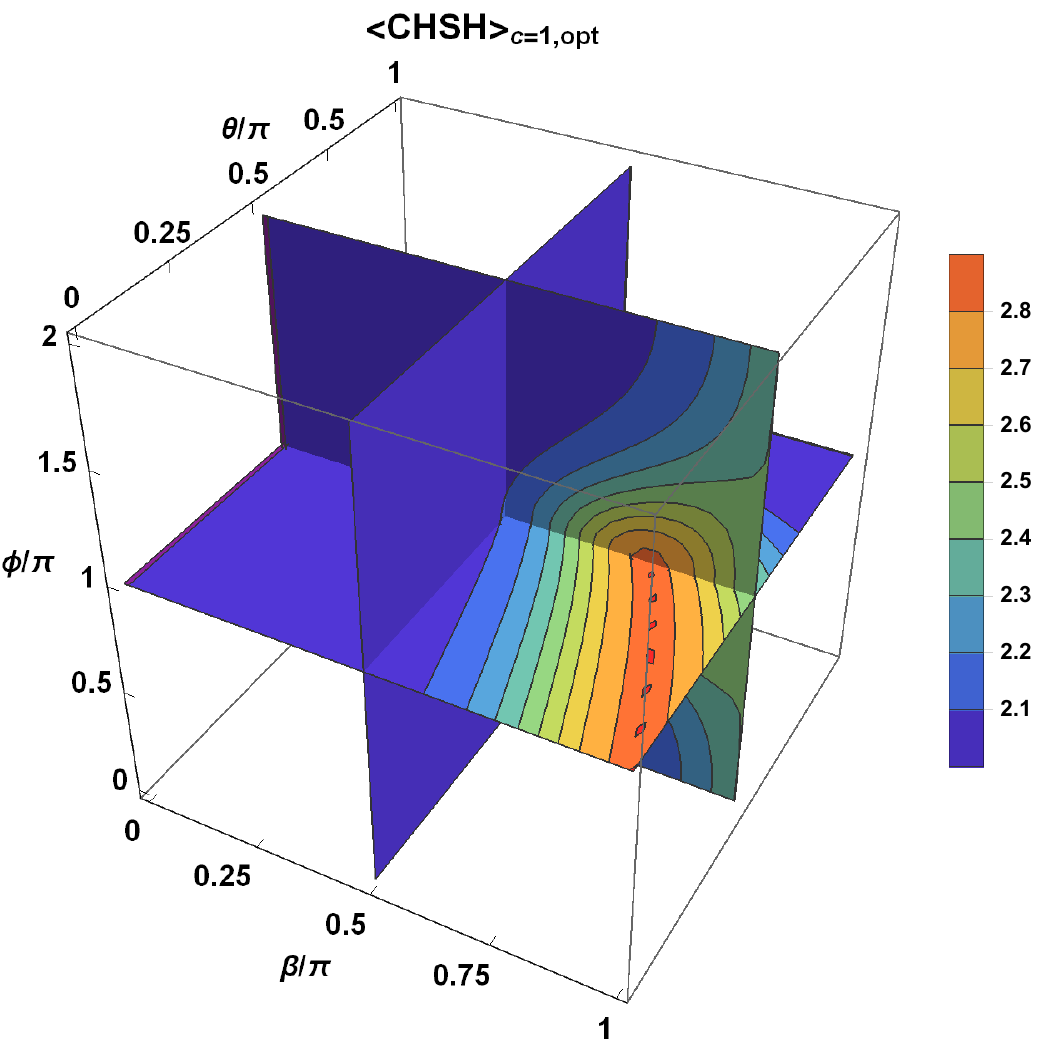}
			\caption{(Colour online) Slice contour plots of the optimal value $\langle {\rm CHSH}\rangle_{c=1,{\rm opt}}$
				as a function of the state parameter $\beta$ and Charlie's  projective measurement orientation angles $(\theta,\, \phi)$ evaluated in the state $\vert \Psi_{3,2}\rangle$ [see (\ref{2meyer})] of the SLOCC class $\{{\cal{D}}_{3,2}\}$. It is evident that the inequality (\ref{chsh_c}) gets violated  in the entire parameter range  $0<\beta\leq \pi$ in the state $\vert \Psi_{3,2}\rangle\in\{{\cal{D}}_{3,2}\}$, when a conditioning over Charlie's measurement outcomes is done. It is also seen that  $\langle{\rm CHSH}\rangle_{c=1,{\rm opt}}$ approaches the allowed maximal value of $2\sqrt{2}$ for a specific range of values of the state parameter $\beta$ and for the orientations $(\theta,\phi)$ of Charlie's measurement.}
		\end{center}
\end{figure}

 In Figure~4 we have presented a slice contour plot of the optimal value  $\langle{\rm CHSH}\rangle_{c=1,{\rm opt}}=2\, \sqrt{(t^{c=1}_1)^2+(t^{c=1}_2)^2}$   evaluated  in the two distinct spinor states $\vert \Psi_{3,2}\rangle$ given by (\ref{2meyer})  as a function of the state parameter $\beta$ and  Charlie's measurement\footnote{Note that  $\langle{\rm CHSH}\rangle_{c=-1,{\rm opt}}=2\, \sqrt{(t^{c=-1}_1)^2+(t^{c=-1}_2)^2}$ is identically equal to $\langle{\rm CHSH}\rangle_{c=1,{\rm opt}}$,  when Charlie changes his measurement orientation $\hat{n}(\theta,\phi)$ to  $-\hat{n}(\theta,\phi)=\hat{n}(\pi-\theta,\pi+\phi).$} settings $\theta$, $\phi$.  It is evident that the one parameter family of states $\vert \Psi_{3,2}\rangle \in\{ {\cal{D}}_{3,2}\}$ violate the conditional CHSH inequality (\ref{chsh_c}) in the entire range of the parameter $0<\beta\leq \pi$. In particular, by conditioning on Charlie's measurement outcome $c=1$ in the $\vert{\rm W}\rangle$ state, obtained by substituting $\beta=\pi$ in $\vert \Psi_{3,2}\rangle$ (see (\ref{W}) and (\ref{2meyer})), it is possible to witness maximum violation $\langle{\rm CHSH}\rangle_{c=1,{\rm opt}}=2\sqrt{2}\approx 2.828$. 

A plot of the optimal value  $\langle {\rm CHSH}\rangle^{\rm con}_{\rm opt}=2\, \sum_{c=\pm 1}\, p(c)\, \sqrt{(t^c_1)^2+(t^c_2)^2}$,
as a function of the state parameter $\beta$ and Charlie's projective measurement settings $\theta$, $\phi$ is given in Figure~5. It is seen that the maximum value of  $\langle{\rm CHSH}\rangle^{\rm con}_{\rm opt}$  in the one-parameter family of states  $\vert \Psi_{3,2}\rangle$ belonging to the SLOCC class $\{{\cal{D}}_{3,2}\}$ can  atmost be equal to 2.55 and does not attain the largest possible value $2\sqrt{2}\approx 2.828$. 
\begin{figure}[h]
	\label{5}
	\begin{center}
		\includegraphics*[width=3in,keepaspectratio]{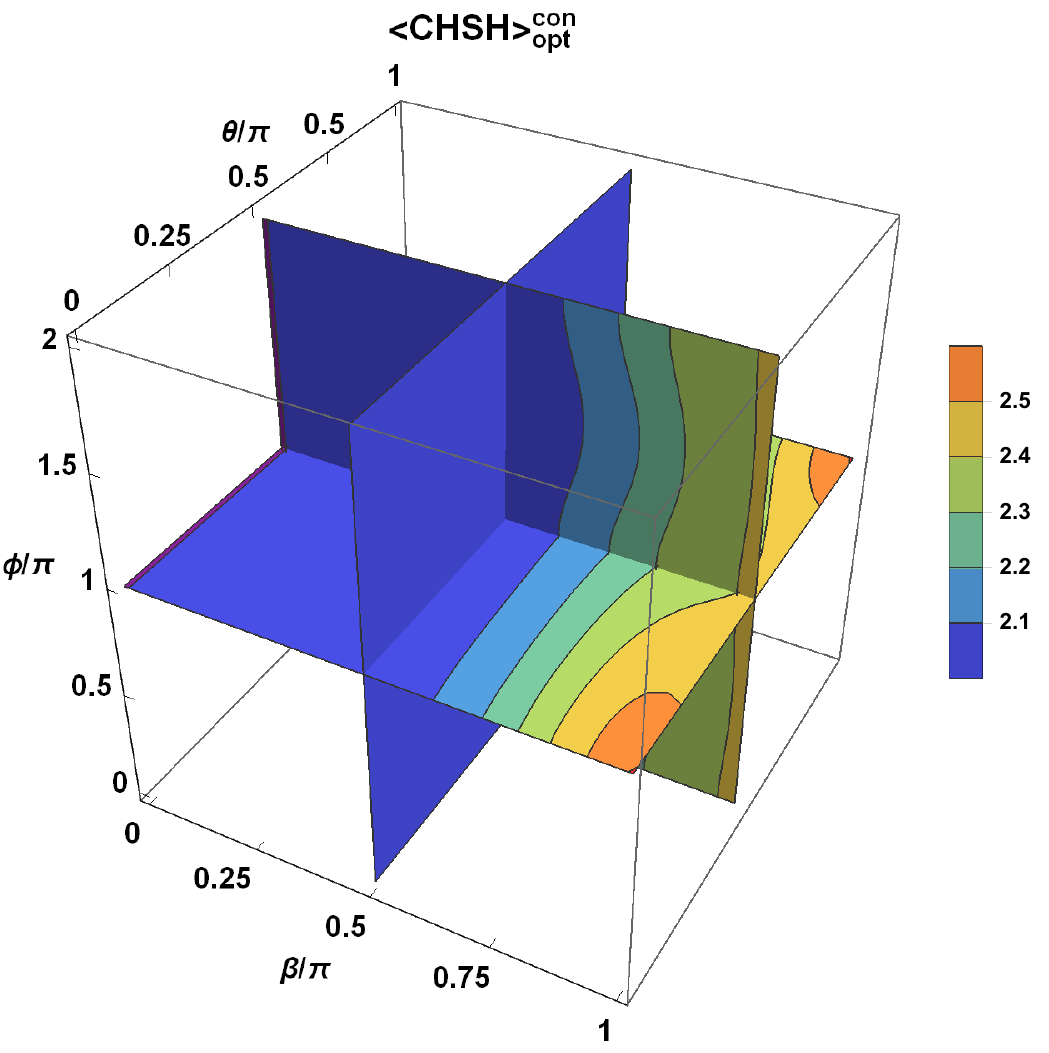}
		\caption{(Colour online) Slice contour plot of the optimal value $\langle{\rm CHSH}\rangle^{\rm con}_{\rm opt}$ in the one parameter family of states $\vert \Psi_{3,2}\rangle$ (see (\ref{2meyer})) of the SLOCC class $\{{\cal{D}}_{3,2}\}$ as a function of the state parameter $\beta$ and Charlie's measurement orientation angles $\theta$, $\phi$. Note that the largest value  $\langle{\rm CHSH}\rangle^{\rm con}_{\rm opt}$ achievable in    the one-parameter family of states  $\vert \Psi_{3,2}\rangle$ belonging to the SLOCC class $\{{\cal{D}}_{3,2}\}$ is ~2.55.}     
	\end{center}
\end{figure} 
    
\subsection{Conditional CHSH nonlocality test on the SLOCC class $\{{\cal{D}}_{3,3}\}$}
We continue to carry out conditional CHSH nonlocality test on the states $\vert \Psi_{3,3}\rangle$   belonging to the three distinct spinor SLOCC class $\{{\cal{D}}_{3,3}\}$. The states defined in (\ref{3meyer}) are characterized by three real parameters. The conditional correlation matrix $T^c$  depends on the state parameters $y, \alpha,\,\beta\,$ and Charlie's measurement orientations $\theta,\phi$. For the sake of illustration we fix $y=1$, $\alpha=0$. With this choice of parameters  the following eigenvalues  $t^c_1,\,t^c_2,t^c_3$ of the conditional correlation matrix $T^c$ are obtained: 	
\begin{eqnarray}
	t^c_{1}&=& 1, \nonumber \\ 
	t^{c}_{2}&=&\frac{1}{\nu_c}\, \left[2\,\left(1+c\,\cos\theta\right)\, \left(1-\, \cos\beta\right)^3\, \left(1 + c\,\cos\theta\, \cos\beta +c\,\cos\phi\,\sin\theta\,\sin\beta\right)\right]^{1/2}\,  \nonumber \\
	&=& \, -t^{c}_{3},
	\end{eqnarray}
	where 
	\begin{eqnarray}
	\nu_c&=&		\cos\phi \sin\theta \sin\beta\, (1+\sin\beta)  
	+ 2 \sin\left(\frac{\beta}{2}\right)\left[2\,c\left(1 + \cos^3\left(\frac{\beta}{2}\right)\right) \right. \nonumber \\ 
	& &\left.+  	\cos\theta  	 \left(1 + 2 \cos^3\left(\frac{\beta}{2}\right) + \cos\beta\right) \right].  
	\end{eqnarray}
	Probability of the measurement outcome $c=\pm 1$  when Charlie performs projective measurement $P_c(\theta,\phi)$ (see (\ref{pcop})) on his qubit is given by 
	\begin{eqnarray}
	p(c)&=&  
	\frac{2+c\, \cos\theta(1+\cos\beta)}{4\,\left[1+ \cos^3\left(\frac{\beta}{2}\right)\right]}  \nonumber   \\ 
		&+& \frac{ 2\,\cos^3\left(\frac{\beta}{2}\right)(1+c\, \cos\theta) + c\, \cos\phi\, \sin\theta\, \sin\beta\, \left(1+  
		\cos\left(\frac{\beta}{2}\right) \right)}{4\,\left[1+ \cos^3\left(\frac{\beta}{2}\right)\right]} \nonumber 
	\end{eqnarray}  
In Figure~6 we present plots of the optimal values  
$\langle{\rm CHSH}\rangle_{c=1,{\rm opt}}=2\sqrt{(t^c_1)^2+(t^c_2)^2}$,  evaluated  in the entangled pure symmetric three-qubit state  $\vert \Psi_{3,3}\rangle$ (see (\ref{3meyer}))  as a function of Charlie's projective measurement orientations $(\theta,\, \phi)$ and  one of the state parameters $0<\beta\leq \pi$ -- for the specific choices $y=1,\, \alpha=0$. (The parameters $y$ and $\alpha$ are chosen in order to witness largest possible violation in this SLOCC class of states). 
\begin{figure}[h]
	\label{6}
		 	\begin{center}
			\includegraphics*[width=3in,keepaspectratio]{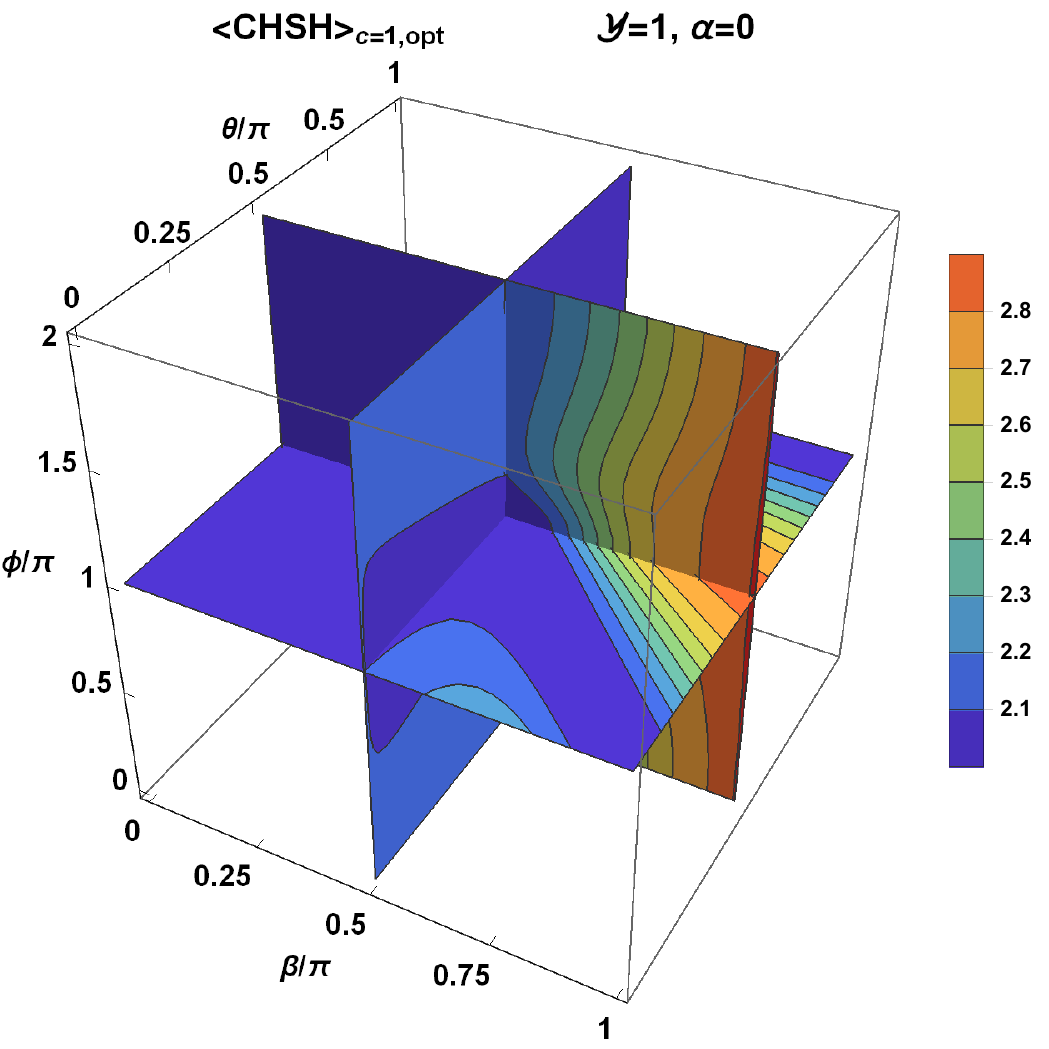}
			\caption{(Colour online) Slice contour plot of  $\langle{\rm CHSH}\rangle_{c=1,{\rm opt}}~=~2\sqrt{(t^{c=1}_1)^2+(t^{c=1}_2)^2}$ for $c=1$, in the  state $\vert \Psi_{3,3}\rangle$ (see (\ref{3meyer})) of the SLOCC class $\{{\cal{D}}_{3,3}\}$ as a function of one of the state parameter $\beta$ and Charlie's projective measurement settings $\theta$, $\phi$. Here we have fixed the values $y=1,\, \alpha=0$ of the other two  parameters characterizing the state. It is seen that the conditional CHSH inequality (\ref{chsh_c}) gets violated upto the allowed maximum value $2\sqrt{2}$ for some specific values of the parameter $\beta$ and Charlie's measurement orientations $\theta$, $\phi$.}     
	\end{center}
	\end{figure}      
It can be seen that for some specific values of $\beta, \theta$ and $\phi$, $\langle{\rm CHSH}\rangle_{c=1,{\rm max}}=2\sqrt{2}$ implying 
maximum violation of conditional CHSH inequality (\ref{chsh_c}).  In particular, GHZ state (obtained by substituting  $\beta=\pi$ in $\vert \Psi_{3,3}\rangle$ of (\ref{3meyer})) violates (\ref{chsh_c}) (as well as (\ref{cdchsh1})) maximally.   
  
We also present contour plot of the optimal conditional CHSH  value (see (\ref{cdopt})) 
\[ 
\langle{\rm CHSH}\rangle^{\rm con}_{\rm opt}=2\, \sum_{c=\pm 1}\, p(c)\, \sqrt{(t^c_1)^2+(t^c_2)^2}
\]  in Figure~7. It is seen that there exist three qubit states belonging to the three distinct spinor SLOCC class $\{{\cal{D}}_{3,3}\}$ which violate the conditional CHSH inequality (\ref{cdchsh1}) all the way up to the  maximum possible value  $2\sqrt{2}$. 

\begin{figure}[h]
	\label{7}
	\begin{center}
		\includegraphics*[width=3in,keepaspectratio]{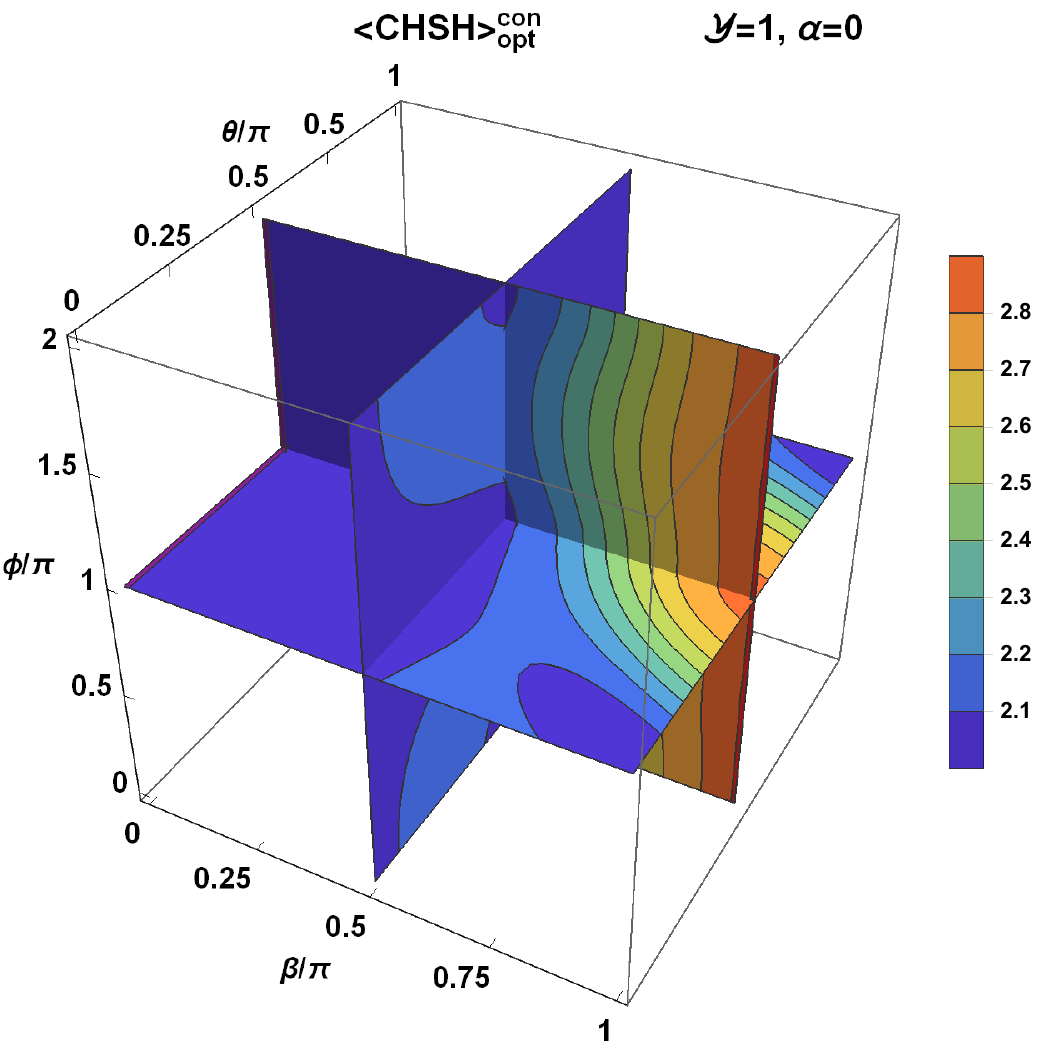}
		\caption{(Colour online) Slice contour plot of $\langle{\rm CHSH}\rangle^{\rm con}_{\rm opt}$ 
			in the three parameter family of  states $\vert \Psi_{3,3}\rangle$  (see (\ref{3meyer})) of the three distinct spinor SLOCC class $\{{\cal{D}}_{3,3}\}$,  as a function of one of the state parameters $\beta$ and Charlie's measurement direction $\theta$ and $\phi$ --  for the specific choice $y=1$, $\alpha=0$.  It is clearly seen that the conditional CHSH value $\langle{\rm CHSH}\rangle^{\rm con}_{\rm opt}$ approaches its allowed maximum value $2\sqrt{2}\approx 2.828$ for specific range of the state parameter $\beta$ and  Chalie's measurement  orientations $\theta,\, \phi$.} 
	\end{center}
\end{figure}
While we have seen from Figure~5 that the maximal value of  $\langle {\rm CHSH}\rangle^{\rm con}_{\rm opt}$ achieved in the one parameter family of states $\vert\Psi_{3,\,2}\rangle$ belonging to the two distinct spinor SLOCC class  $\{{\cal D}_{3,2}\}$  is  $~2.55$, 
from Figure~7 it is evident that the largest  value of $\langle {\rm CHSH}\rangle^{\rm con}_{\rm opt}$ for the states $\vert\Psi_{3,\,3}\rangle$ of the class $\{{\cal D}_{3,3}\}$ goes up to the largest possible value $2\sqrt{2}$. This prompts us to quantify the nonlocality of entangled  three-qubit symmetric states by defining the quantity ${\cal Q}$
\be
		{\cal Q}={\rm max}\left\{0,\, \frac{\langle {\rm CHSH}\rangle^{\rm con}_{\rm opt}-2}{2\sqrt{2}-2}\right\}
		\ee  
		which increases from 0 to 1  as $\langle {\rm CHSH}\rangle^{\rm con}_{\rm opt}$ changes from $2$ to $2\sqrt{2}.$ Based on numerical evaluations it is found that  the nonlocality quantifier ${\cal Q}$ does not vanish in the entire range of  parameters of the three-qubit states $\vert\Psi_{3,\,2}\rangle, \ \vert\Psi_{3,\,3}\rangle$. While it is seen  that a maximum of  ${\cal Q}\approx 0.66$ is obtained for $\beta=\pi$ and for Charlie's measurement orientations $\theta=0,\,\pi,\  \phi=0$ in $\vert\Psi_{3,\,2}\rangle\in \{{\cal D}_{3,2}\}$, one obtains ${\cal Q}=1$ when  $y=1,\beta=\pi,\,\alpha=0$,  for measurement angles $\theta=\pi/2,\, \phi=0$ in  $\vert\Psi_{3,\,3}\rangle\in \{{\cal D}_{3,3}\}.$  In other words, the conditional CHSH nonlocality test (see (\ref{cdchsh1}), (\ref{cdopt})), quantified via ${\cal Q}$ serves as a witness for discriminating entangled pure symmetric three-qubit states belonging to the  inequivalent SLOCC classes $\{{\cal{D}}_{3,2}\},\ \{{\cal{D}}_{3,3}\}$. We have plotted ${\cal Q}$ in Figures~8 and 9 for both the SLOCC classes $\{{\cal D}_{3,2}\}$, $\{{\cal D}_{3,3}\}$ (in Figure~9, we have fixed the parameters $y=1,\ \alpha=0$ in $\vert\Psi_{3,\,3}\rangle$ and orientation angle $\phi$ of Charlie's measurement is chosen to be zero for the purpose of illustration). 
		
     
     \begin{figure}[h]
     	\label{8}
     	\begin{center}
     		\includegraphics*[width=3in,keepaspectratio]{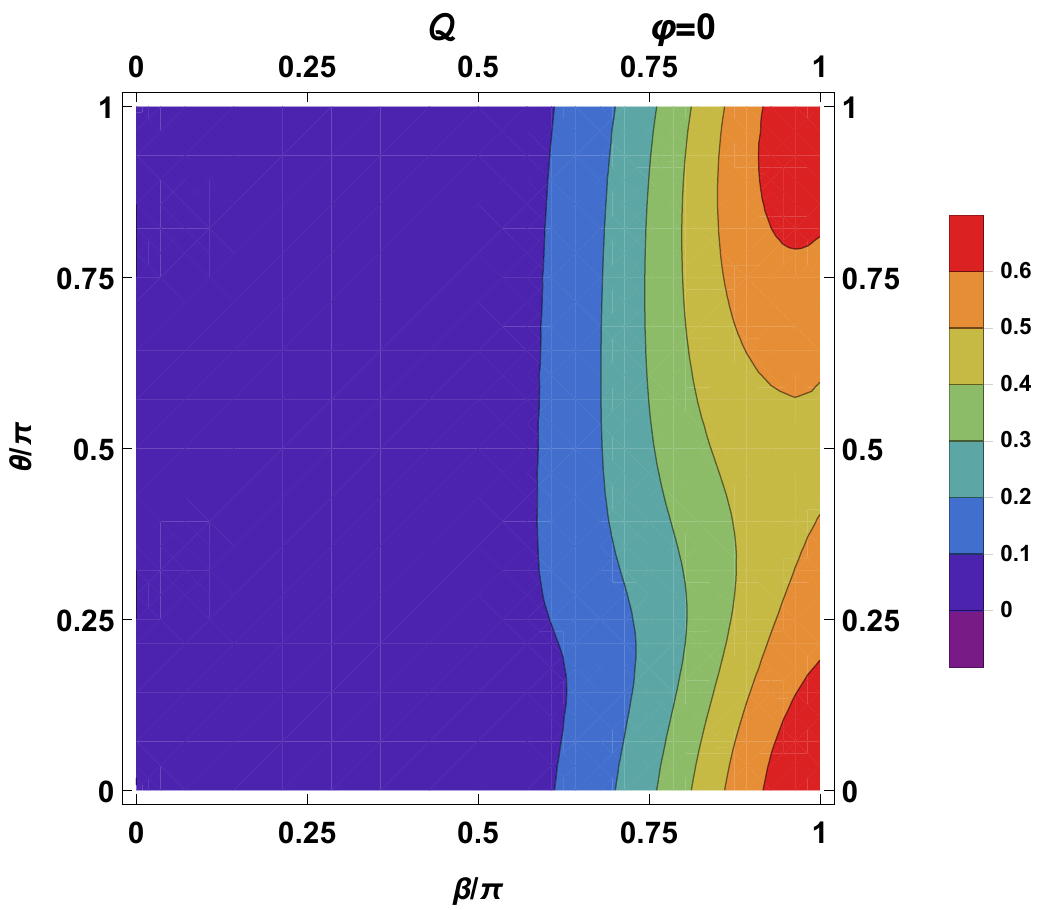}
     		\caption{(Colour online) Plot of the nonlocality quantifier ${\cal Q}$ 
     			evaluated in the   state $\vert \Psi_{3,2}\rangle$ of the SLOCC class $\{{\cal{D}}_{3,2}\}$,  as a function of one of the state parameters $\beta$ and Charlie's measurement orientations  $\theta$. Here we have fixed $\phi=0$.  It is seen that ${\cal Q}$ is non-zero in the entire range of the state parameter $0<\beta\leq \pi$ and reaches a maximum value $\approx 0.66$, when $\beta=\pi$ (which corresponds to the W state), $\theta=0,\pi$.} 
     	\end{center}
     \end{figure}
     \begin{figure}[h]
     	\label{9}
     	\begin{center}
     		\includegraphics*[width=3in,keepaspectratio]{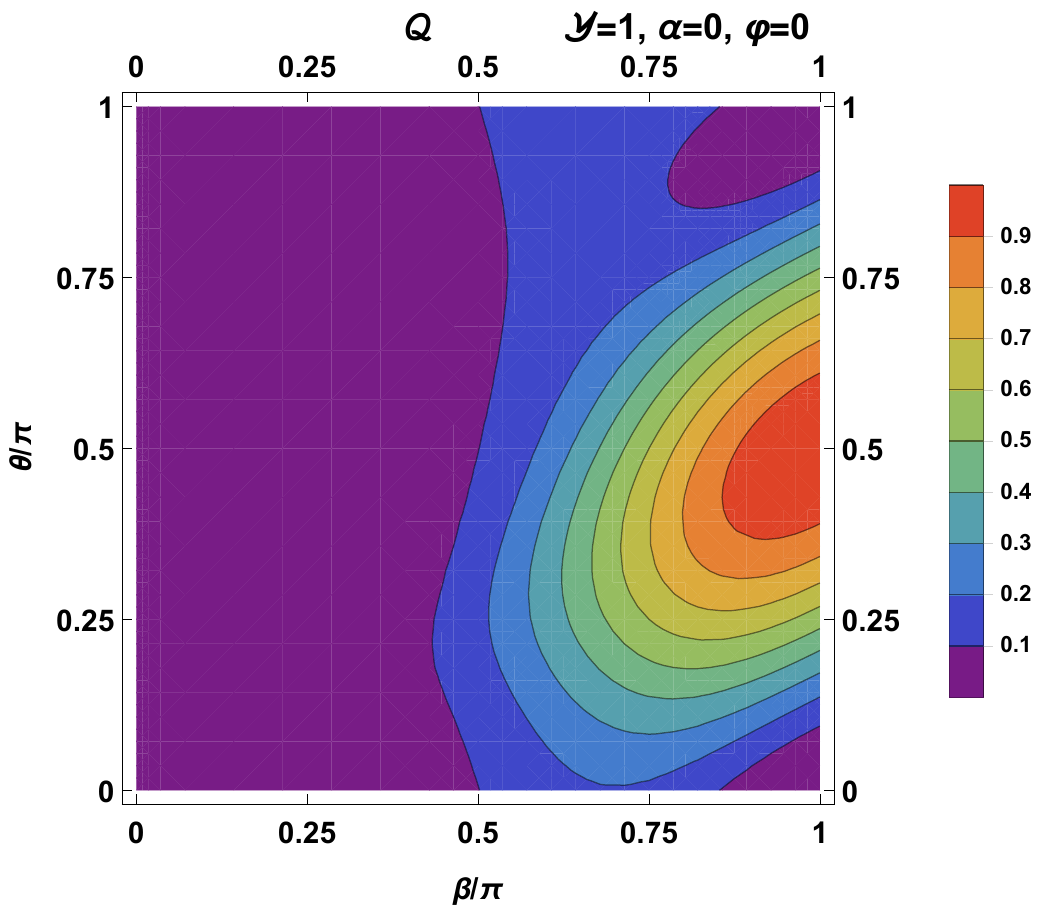}
     		\caption{(Colour online) Plot of the nonlocality quantifier ${\cal Q}$ 
     			evaluated in the   state $\vert \Psi_{3,3}\rangle\in\{{\cal{D}}_{3,3}\}$,  as a function of  $\beta$ and Charlie's measurement orientations $\theta$;  we have fixed the parameters $y=1,\alpha=0$ and chosen the measurement angle $\phi=0$. We find that  ${\cal Q}$ is non-zero for the entire SLOCC class $\{{\cal{D}}_{3,3}\}$ and the maximum value ${\cal Q}=1$ is obtained when the parameter $\beta=\pi$ (which corresponds to the GHZ state) for the measurement angle $\theta=\pi/2$.} 
     	\end{center}
     \end{figure}       
Recall that  entanglement of  the three qubit states belonging to the three distinct spinor SLOCC class $\{{\cal{D}}_{3,3}\}$ (see (\ref{3meyer})) is characterized  by  the {\em three-tangle} $\tau$ and the {\em pairwise concurrence} $C$, expressed in terms of the state parameters $0<y\leq 1,0\leq \alpha\leq 2\pi, 0<\beta\leq \pi$ as~\cite{meyer}  
\begin{eqnarray}
\label{tangle}
\tau=\frac{2y \sin^3(\beta/2)}{1+y^2+2y\,\cos^3(\beta/2)\cos\alpha} \\ 
\label{con33}
C=\frac{y \sin(\beta)\, \sin(\beta/2)}{1+y^2+2y\,\cos^3(\beta/2)\cos\alpha}.
\end{eqnarray}  
However, the three-tangle vanishes for the three qubit states belonging to the two distinct spinor class $\{{\cal{D}}_{3,2}\}$ and  the  pairwise concurrence is given by~\cite{meyer}   
\begin{eqnarray}
\label{con32}
C=\frac{2-2 \cos^2(\beta/2)}{3+6\cos^2(\beta/2)}.
\end{eqnarray}  
   
Note that the nonlocality quantifier ${\cal Q}$ of the class $\{{\cal{D}}_{3,3}\}$ assumes its largest possible value in the GHZ state,  characterized by the state parameters $y=1,\, \beta=\pi,\alpha=0$, (see (\ref{3meyer})) for which the three-tangle attains its maximum value   $\tau=1$ and the pairwise concurrence $C$ vanishes (see (\ref{tangle}) and (\ref{con33})). This indicates that the three-tangle $\tau$ plays a significant role in characterizing the non-locality of three qubit states of the  SLOCC family $\{{\cal{D}}_{3,3}\}$  under the conditional-CHSH test. On the other hand, the  the non-locality quantifier ${\cal Q}$ associated with the SLOCC class $\{{\cal{D}}_{3,2}\}$  attains its maximum value in the W state, governed by the state parameter $\beta=\pi$ for which the  pairwise concurrence $C$ (see  (\ref{con32}))  approaches its highest possible value.  In other words, the conditional CHSH test indicates that genuine tripartite non-local correlations characterized by non-zero tangle are stronger than those characterized by pairwise concurrence. In the next section we proceed to explore non-locality of pure permutation symmetric three-qubit states using tight Bell inequalities in the (3,2,2) scenario.

\section{Nonlocality of pure symmetric three-qubit states via tight Bell inequalities in the (3,2,2) scenario}
In this section we discuss maximum nonlocality of pure entangled symmetric three-qubit states using the tight Bell inequalities in the (3,2,2) i.e., three-party, two-setting, two-outcome   scenario~\cite{pit01,silwa,cab1}. Pitowsky and Svozil~\cite{pit01}, Sliwa~\cite{silwa} had identified that there are exactly 46 inequivalent classes of tight Bell inequalities in the (3,2,2) scenario, which get maximally violated by entangled three-qubit states, when suitable local measurements  are carried out. These 46 classes of tight Bell inequalities and their local maximum value have been listed in Table~1 of Ref.~\cite{cab1}. It has been shown~\cite{LMas} that maximum violations of the tight Bell inequalities are attained by three-qubit pure states when local projective measurements are employed. In Ref.~\cite{cab1}, explicit structure  of the set of all entangled pure three-qubit states  along with the  optimal  local observables and the corresponding quantum maximum value achievable in these states has been tabulated. We observe that  {\em six} classes of these tight Bell inequalities i.e., 2, 5, 22, 26, 33 and 39 (see Table~2 of Ref.~\cite{cab1}) 
are maximally violated by pure permutation symmetric three-qubit states~\footnote{The three-qubit  states $\vert\psi^{(k)}_{\rm ABC}\rangle, k=2,5,22,26,33,39$ are the  ones which exhibit identical pairwise concurrences $C_{\rm AB}=C_{\rm BC}=C_{\rm AC}$ (see Table~VI of Ref.~\cite{cab1}).  These states are found to be local unitary equivalent to permutation symmetric states $\vert\psi^{(k)}_{\rm sym}\rangle$.}.  We have listed these six classes of tight Bell inequalities, the optimal local dichotomic measurement settings  $A_i,\,B_i,\,C_i,i=1,2$ of Alice, Bob and Charlie,  the three-qubit pure state $\vert\psi^{(k)}_{\rm ABC}\rangle,\ k=2, 5, 22, 26, 33, 39$  in which maximum violation is witnessed are given  in Table~1. 

 \begin{table}[htbp]
 		\label{I}
			\caption{A list of six different classes  of tight Bell inequalities, labeled by 2, 5, 22, 26, 33 and 39,  [see Tables~I,~IV and ~V  of Ref.~\cite{cab1} for details on the 46 classes of tight Bell inequalities of the (3,2,2) scenario], corresponding local dichotomic observables  $A_i,\, B_i,\, C_i,i=1,2$ of Alice, Bob, Charlie, pure three-qubit states  $\vert\psi^{(k)}_{\rm ABC}\rangle$ which maximally violate the inequalities, \&  the  maximum quantum violation.}  	  
			\begin{tabular}{|c|c|c|c|}
				\hline 
				Class &  Tight Bell inequality   & State  & Quantum  \\
				&   &  &  maximum  \\ 
				\hline
				2	& \begin{tabular}{c}  $\left\langle   (A_1\, B_1 + A_2\,B_2)\,C_1 + (A_2\, B_1 -A_1\,B_2)\,C_2\right\rangle \leq 2,$ \\ 
					\\
					$A_1=\sigma_3\otimes I\otimes I,\,A_2= \sigma_1\otimes I\otimes I$   \\ 
					$B_1= I\otimes\sigma_3\otimes I,\, B_2=I\otimes \sigma_1\otimes I$   \\
					$C_1= I\otimes I \otimes \sigma_3,\, C_2=I\otimes I\otimes \sigma_1$  \\ 
				\end{tabular} &  
				\begin{tabular}{l} 		 $\vert \psi_{\rm ABC}^{(2)}\rangle=\frac{1}{\sqrt{2}}\left(\vert \tilde{+} \tilde{-}\tilde{-}\rangle + \vert \tilde{-}\tilde{+}\tilde{+}\rangle  \right)$ \\ 
					\\ 
					$\vert \tilde{\pm}\rangle=\frac{1}{\sqrt{2}}(\vert 0\rangle\pm i\vert 1\rangle)$	
				\end{tabular} & 4 \\ 
				\hline	
				5 & 	\begin{tabular}{c}	$\left\langle (A_1+B_1+C_1)+B_1\,C_2 + B_2\,(C_1-C_2)  \right.$  \\
					$+ A_1\,(B_2+C_2-B_1\,C_1-B_1\,C_2- B_2\,C_1)$ \\   
					$\left. + A_2\,(B_1+C_1-B_2-C_2-B_1\,C_1+ B_2\,C_2)\right\rangle\leq 3$  \\   
					\\
					$A_1=o^{(5)}_1\otimes I\otimes I,\,A_2=\sigma_1\otimes I \otimes I$   \\  
					$B_1= I\otimes o^{(5)}_1\otimes I,\, B_2=I\otimes \sigma_1\otimes I$  \\
					$C_1= I\otimes I \otimes o^{(5)}_1,\, C_2=I\otimes I\otimes \sigma_1$  \\   
					$o^{(5)}_1=\sqrt{-2+\sqrt{5}}\,(2\,\sigma_3+\sigma_1)$  \\ 
				\end{tabular} &  \begin{tabular}{c}
					$\vert \psi_{\rm ABC}^{(5)}\rangle=a^{(5)}\,\vert\bar{+}\bar{+}\bar{+}\rangle+ b^{(5)}\, \vert\bar{-}\bar{-}\bar{-}\rangle$ \\ 	 
					\hskip 0.5in$ +c^{(5)}\,  \vert\bar{+}\bar{+}\bar{-}\rangle+d^{(5)}\,  
					\vert\bar{+}\bar{-}\bar{+}\rangle+e^{(5)}\,  
					\vert\bar{-}\bar{+}\bar{+}\rangle$   \\  \\ 
					$a^{(5)}=-\left(-\frac{3}{2}+\frac{7}{2\,\sqrt{5}}\right)^{1/2},\, b^{(5)}=-\left(1-\frac{2}{\sqrt{5}}\right)^{1/2}$\\ 
					$c^{(5)}=d^{(5)}=e^{(5)}=\left(\frac{1}{2}-\frac{1}{2\,\sqrt{5}}\right)^{1/2}$ \\ \\
					$\vert \bar{\pm}\rangle=\frac{1}{\sqrt{2}}(-\vert 0\rangle\pm \vert 1\rangle)$  \\	 
				\end{tabular} &  $8\sqrt{5}-13$ \\ 
				\hline
				22		&  \begin{tabular}{c}   
					$\left\langle (A_1+B_1+C_1)+(A_2+B_2+C_2)+B_1\,C_1-B_2\,C_2  \right.$ \\
					$+A_1\,(B_1+C_1- 2\, B_1\,C_1-B_2\,C_1-B_1\,C_2+B_2\,C_2)$ \\ 
					$\left. -A_2(B_2+C_2+B_1\,C_1-B_1\,C_2-B_2\,C_1) \right\rangle\leq 4 $ \\	 \\	
					$A_1=o^{(22)}_1\otimes I\otimes I,\,A_2=o^{(22)}_2\otimes I\otimes I$\\   
					$B_1= I\otimes o^{(22)}_1\otimes I,\, B_2=I\otimes o^{(22)}_2\otimes I$\\ 
					$C_1= I\otimes I \otimes o^{(22)}_1,\, C_2=I\otimes I\otimes o^{(22)}_2$\\ 
					$o^{(22)}_1=-0.25333\,\sigma_1+0.96738\,\sigma_3$ \\
					$o^{(22)}_2=0.99937\,\sigma_1+0.03540\,\sigma_3$ \\
				\end{tabular}             &  
				\begin{tabular}{c}      $\vert \psi_{\rm ABC}^{(22)}\rangle=a^{(22)}\,\vert000\rangle+ b^{(22)}\, \vert 111\rangle$  \\  
					$\hskip 0.5in	+c^{(22)}\,  \vert 001\rangle+d^{(5)}\,  
					\vert 100\rangle+\rangle+e^{(5)}\,  
					\vert 010 \rangle$    \\  \\ 
					$a^{(22)}=0.161337,\, b^{(22)}=-0.664411$ 	 \\	
					$c^{(22)}=d^{(22)}=e^{(22)}=0.421319$ 	\\
				\end{tabular}  & 6.19794 \\
				\hline
				26 & \begin{tabular}{c} $\left\langle (A_1+B_1+C_1)+A_1(B_1+C_1-B_1\,C_1+2\, B_2\,C_2) \right.$   \\
					$+ 2A_2\,(B_2+C_2-B_1\,C_2- B_2\,C_1)$ \\
					$\left. +B_1\,C_1 -2\,B_2\,C_2\right\rangle \leq 5$ \\  \\
					$A_1=\sigma_3\otimes I\otimes I,\,A_2= \sigma_1\otimes I\otimes I$   \\ 
					$B_1= I\otimes\sigma_3\otimes I,\, B_2=I\otimes \sigma_1\otimes I$ \\
					$C_1= I\otimes I \otimes \sigma_3,\, C_2=I\otimes I\otimes \sigma_1$  \\
				\end{tabular}
				& \begin{tabular}{c}
					$\vert \psi_{\rm ABC}^{(26)}\rangle=\frac{1}{\sqrt{6}}\,(	\vert 001\rangle+\vert 010\rangle)-\vert 100\rangle)
					+\frac{1}{\sqrt{2}} \vert 111\rangle$ \\ 
				\end{tabular} &   $1+ 4\sqrt{3}$ \\
				\hline 
				33 & \begin{tabular}{c}  $\left\langle (A_1+B_1+C_1)+ (A_2+B_2+C_2)  -B_1\,C_2-B_2\,C_1\right.$ \\
					$ -A_2\,(B_1+C_1 - B_1\,C_2 - 2\, B_1\,C_1 - B_2\,C_1  + 3\, B_2\,C_2) $  \\ 
					$ \left.    - A_1\,(B_2 + C_2  - 2 \, B_1\,C_2- 2\, B_2\,C_1 -  B_2\, C_2)\right\rangle \leq 6$ \\ \\
					$A_1=o^{(33)}_1\otimes I\otimes I,\,A_2= o^{(33)}_2\otimes I\otimes I$ \\
					$B_1= I\otimes o^{(33)}_1\otimes I,\, B_2=I\otimes o^{(33)}_2\otimes I$  \\
					$C_1= I\otimes I \otimes o^{(33)}_1,\, C_2=I\otimes I\otimes o^{(33)}_2$ \\
					$o^{(33)}_1=0.48263\,\sigma_1+0.87582\,\sigma_3$\\ 
					$o^{(33)}_2=0.92087\,\sigma_1-0.38987\,\sigma_3$  \\
				\end{tabular} 
				& \begin{tabular}{c}
					$\vert \psi_{\rm ABC}^{(33)}\rangle=a^{(33)}\,\vert000\rangle+ b^{(33)}\, \vert 111\rangle $     \\ 
					$\hskip 0.5in	+c^{(33)}\,  \vert 001\rangle+d^{(33)}\,  
					\vert 100\rangle+e^{(33)}\,  
					\vert 010 \rangle$  \\   \\ 
					$a^{(33)}=-0.024223,\, b^{(33)}=-0.621262$,    \\
					$c^{(33)}=d^{(33)}=e^{(33)}=0.452197$	\\
				\end{tabular} 
				& 9.78988 \\
				\hline 
				39 &  \begin{tabular}{c}  $\left\langle  2\, (A_1+B_1+C_1)+B_1\,(C_2-C_1) + B_2(C_1+C_2) \right.$  \\ 
					$- A_1\,(B_1+C_1-B_2-C_2)+A_2\,(B_1+C_1+ B_2+C_2)$ \\  
					$+ A_1(2\,B_1\,C_1-B_1\,C_2-\,B_2\,C_1-2\,B_2\,C_2)   $  \\ 
					$\left.  - A_2\,(B_1\,C_1+2\,B_1\,C_2+2\,B_2\,C_1-B_2\,C_2)\right\rangle\leq 6$ \\ \\
					$A_1=o^{(39)}_1\otimes I\otimes I,\,A_2= o^{(39)}_2\otimes I\otimes I$   \\
					$B_1= I\otimes o^{(39)}_1\otimes I,\, B_2=I\otimes o^{(39)}_2\otimes I$\\ 
					$C_1= I\otimes I \otimes o^{(39)}_1,\, C_2=I\otimes I\otimes o^{(39)}_2$\\
					$o^{(39)}_1=-0.04834\,\sigma_1-0.99883\,\sigma_3$  \\
					$o^{(39)}_2=-0.99683,\sigma_1+0.07953\,\sigma_3$  \\
				\end{tabular} 
				& 
				\begin{tabular}{c}
					$\vert \psi_{\rm ABC}^{(39)}\rangle=a^{(39)}\,\vert000\rangle+ b^{(39)}\, 111\rangle$\\ 
					$+c^{(39)}\,  
					\vert 001\rangle+d^{(39)}\,\vert 100\rangle+e^{(39)}\,\vert 010\rangle$\\  \\
					$a^{(39)}=0.177347,\, b^{(39)}=-0.386311,$ \\  
					$ c^{(39)}=d^{(39)}=e^{(39)}=0.522594$ \\	
				\end{tabular} & 9.32530 \\
				\hline 		 
		\end{tabular}
	\end{table}

 It is readily identified that the states $\psi^{(k)}_{\rm ABC}\rangle,\, k=2, 5, 22, 26, 33, 39$ are local unitary equivalent to permutation symmetric states $\vert\psi^{(k)}_{\rm sym}\rangle$: 
\begin{eqnarray*}
\label{aazz1}
{\cal U}^{(2)}\, \vert\psi^{(2)}_{\rm ABC}\rangle&=& \vert\psi^{(2)}_{\rm sym}\rangle= \vert {\rm GHZ}\rangle \nonumber \\
{\cal U}^{(5)}\,\vert\psi^{(5)}_{\rm ABC}\rangle&=& \vert\psi^{(5)}_{\rm sym}\rangle \\ &=& a^{(5)}\,\vert 000\rangle + b^{(5)}\,\vert 111\rangle +\sqrt{3}\,c^{(5)}\,\vert {\rm W}\rangle \nonumber \\ 
{\cal U}^{(0)}\vert\psi^{(22)}_{\rm ABC}\rangle&=& \vert\psi^{(22)}_{\rm sym}\rangle\\ &=&a^{(22)}\,\vert 000\rangle + b^{(22)}\,\vert 111\rangle +\sqrt{3}\,c^{(22)}\,\vert {\rm W}\rangle \nonumber \\ 
{\cal U}^{(26)}\,\vert\psi^{(26)}_{\rm ABC}\rangle&=& \vert\psi^{(26)}_{\rm sym}\rangle = -\frac{1}{\sqrt{2}}(\vert 111\rangle + \vert {\rm W}\rangle) \nonumber  \\
{\cal U}^{(0)}\vert\psi^{(33)}_{\rm ABC}\rangle&=&\vert\psi^{(33)}_{\rm sym}\rangle \\ &=& a^{(33)}\,\vert 000\rangle + b^{(33)}\,\vert 111\rangle +\sqrt{3}\,c^{(33)}\,\vert {\rm W}\rangle\nonumber \\ 
{\cal U}^{(0)}\,\vert\psi^{(39)}_{\rm ABC}\rangle&=&\vert\psi^{(39)}_{\rm sym}\rangle \\ &=& a^{(39)}\,\vert 000\rangle + b^{(39)}\,\vert 111\rangle +\sqrt{3}\,c^{(39)}\,\vert {\rm W}\rangle
\end{eqnarray*} 
where 
\begin{eqnarray*}
\label{aazz2}
	{\cal U}^{(0)} &=& I\otimes I\otimes I \\
{\cal U}^{(2)}&=& U_{-}^{(2)}  \otimes U_{+}^{(2)}\otimes U_{+}^{(2)};\ \ U_{\pm}^{(2)}=\frac{1}{\sqrt{2}}\ba{cc} 1 & \pm i \\ 1 & \mp i \ea\\
{\cal U}^{(5)}&=& U^{(5)}\otimes U^{(5)}\otimes U^{(5)}, \ U^{(5)}=\frac{1}{\sqrt{2}}\ba{cc} -1 &  1 \\ -1 &   -1 \ea\\
{\cal U}^{(26)}&=& \sigma_3\otimes I\otimes  I 
\end{eqnarray*}

In  Table~2 we
 have displayed the explicit forms of the distinct constituent spinors of the pure three-qubit symmetric  states~\footnote{We determine the explicit form $\vert \beta\rangle=\cos(\beta/2)\vert 0\rangle+\sin(\beta/2)\vert 1\rangle$ of the constituent qubit states of the  three-qubit pure symmetric states $\vert\psi^{(k)}_{\rm sym}\rangle$, $k=2,\,5,\,22,\,33,\,39$ by solving the Majorana polynomial equation~\cite{aru}  $b^{(k)}~+~3\,c^{(k)}  z^2-a^{(k)}\, z^3  =0,$ where $z=\tan(\beta/2)$. The Majorana polynomial equation associated with the state $\vert\psi^{(26)}_{\rm sym}\rangle$ reduces to  $z^{-1}\,(-1+\sqrt{3}\,z^2)=0$.} $\vert \psi^{(k)}_{\rm sym}\rangle$, $k=2,\, 5,\, 22,\, 26,\, 33,\, 39$. All these states belong to the three distinct spinor SLOCC class $\{{\cal{D}}_{3,3}\}.$  None of the  tight Bell inequalities of the (3,2,2) scenario are maximally violated by permutations symmetric three-qubit pure states of the two distinct spinor class $\{{\cal{D}}_{3,2}\}$. On the other hand, there are {\emph six} three-qubit pure symmetric states $\vert \psi^{(k)}_{\rm sym}\rangle$  $k=2,\,5,\,22,\,26,\,33,\,39$ of the SLOCC class $\{{\cal D}_{3,3}\}$, which maximally  violate six classes of  tight Bell inequalities (see Table~1).  Violation of the tight Bell inequality `26'  by the three-qubit state $\vert\psi^{(26)}_{\rm ABC}\rangle$ has been experimentally verified recently~\cite{cab2}.   

\begin{table}
 \caption{Three distinct spinors constituting the pure three-qubit  symmetric states $\vert\psi^{(k)}_{\rm sym}\rangle$.  The states $\vert\psi^{(k)}_{\rm sym}\rangle$ belong to the SLOCC class $\{{\cal D}_{3,3}\}$ of three distinct spinors and they maximally violate the tight Bell inequalities $k=2,\,5,\,22,\,26,\,33,\,39$ of the (3,2,2) scenario~\cite{cab1}).}
\hskip 1in \begin{tabular}{|c|c|}
			\hline
 		Class &	Three-qubit symmetric states $\vert\psi^{(k)}_{\rm sym}\rangle$ \& \\ 
		& the associated       
		 constituent spinors \\ 
			\hline 
	2 & 		$\vert \psi^{(2)}_{\rm sym}\rangle ={\cal N}^{(2)}\, \sum_P \hat{P}\{ \vert\, \phi\rangle_1 \,  \otimes \vert\, \phi\rangle_2  \otimes \vert \phi\rangle_3 \} =\vert {\rm GHZ}\rangle$ \\ 
	&	$\vert\phi\rangle_p=\frac{1}{\sqrt{2}}(\vert 0\rangle+\omega^p\,\vert 1\rangle), p=1,2,3$      
			where $\omega^3=1$  \\
\hline 
 & 	$\vert \psi^{(5)}_{\rm sym}\rangle ={\cal N}^{(5)}\, \sum_P \hat{P}\{ \vert\, \beta\rangle_1 \,  \otimes \vert\, \beta\rangle_2  \otimes \vert \beta\rangle_3 \}$   \\ 
5 & 	 
 $\vert\beta\rangle_1=0.916\, \vert 0\rangle +0.402\,\vert 1\rangle,$   
	$\vert\beta\rangle_2=0.904\, \vert 0\rangle -0.427\,\vert 1\rangle$  \\
& 	 $\vert\beta\rangle_3=0.161\, \vert 0\rangle -0.987\,\vert 1\rangle$   \\ 
	\hline
 & 	$\vert \psi^{(22)}_{\rm sym}\rangle ={\cal N}^{(22)}\, \sum_P \hat{P}\{ \vert\, \beta\rangle_1 \,  \otimes \vert\, \beta\rangle_2  \otimes \vert \beta\rangle_3 \}$   \\   
22 & 	$\vert\beta\rangle_1=\,0.795 \vert 0\rangle +0.607\,\vert 1\rangle,$    
	$\vert\beta\rangle_3=\,0.821 \vert 0\rangle -0.571  \,\vert 1\rangle$   \\
& 	$\vert\beta\rangle_2=\,0.128 \vert 0\rangle + 0.992\,\vert 1\rangle$    \\
	\hline  
 & 	 $\vert \psi^{(26)}_{\rm sym}\rangle ={\cal N}^{(26)}\, \sum_P \hat{P}\{ \vert\, \beta\rangle_1 \,  \otimes \vert\, \beta\rangle_2  \otimes \vert \beta\rangle_3 \}$   \\ 
26 & 	$\vert\beta\rangle_1=\,0.796 \vert 0\rangle +0.605\,\vert 1\rangle$, 
	$\vert\beta\rangle_2=\,0.796 \vert 0\rangle - 0.605\,\vert 1\rangle$    \\
& 	$\vert\beta\rangle_3=\vert 1\rangle$ \\  
	\hline 
  & 	$\vert \psi^{(33)}_{\rm sym}\rangle ={\cal N}^{(33)}\, \sum_P \hat{P}\{ \vert\, \beta\rangle_1 \,  \otimes \vert\, \beta\rangle_2  \otimes \vert \beta\rangle_3 \}$   \\ 
33 & 	$\vert\beta\rangle_1=0.830\, \vert 0\rangle + 0.558  \,\vert 1\rangle$,    
	$\vert\beta\rangle_2= 0.826 \, \vert 0\rangle - 0.562\,\vert 1\rangle$  \\
& $\vert\beta\rangle_3=0.0018    \, \vert 0\rangle-0.998\,\vert 1\rangle$  \\ 
	\hline  
 & $\vert \psi^{(39)}_{\rm sym}\rangle ={\cal N}^{(39)}\, \sum_P \hat{P}\{ \vert\, \beta\rangle_1 \,  \otimes \vert\, \beta\rangle_2  \otimes \vert \beta\rangle_3 \}$   \\ 
39 & $\vert\beta\rangle_1=0.890\, \vert 0\rangle+0.455\,\vert 1\rangle$,    
$\vert\beta\rangle_2=0.900  \, \vert 0\rangle - 0.435\,\vert 1\rangle$   \\
& $\vert\beta\rangle_3=0.113\, \vert 0\rangle + 0.994 \,\vert 1\rangle$  \\ 
\hline	
\end{tabular}
\end{table}

\section{Summary} Different forms of Bell inequalities have been proposed for  studying nonlocality of permutation symmetric multiqubit states~\cite{lewenstein1,lewenstein2}. In this work, we have focused on the  CHSH~\cite{chsh} and the conditional CHSH inequalities~\cite{cavalcanti,rchaves} to investigate nonlocal features of pure three-qubit symmetric states. These states admit an  elegant parametrization for different SLOCC classes based on Majorana geometric representation~\cite{meyer}. Making use of the explicit parametrization we prove that  pairs of qubits  drawn from a pure entangled three-qubit symmetric state (shared by Alice, Bob and Charlie) are  CHSH local. In fact this property can be attributed to  monogamy of CHSH-nonlocality~\cite{Qin15,Hall17}. Our explicit verification $\langle {\rm CHSH}\rangle_{\rm opt}\leq 2$  by two-qubit correlations  of  the entire class of entangled pure symmetric three qubit states thus upholds CHSH monogamy property. 

Continuing further we have shown that conditional CHSH inequalities~\cite{cavalcanti,rchaves} are useful in activating 
nonlocality hidden in two-qubit correlations recorded by Alice and Bob, when they are conditioned by the dichotomic outcomes of Charlie's measurement on his qubit. We have explicitly demonstrated that the permutation symmetric three qubit states belonging to both the two and three distinct spinor SLOCC classes  $\{{\cal D}_{3,2}\}$, $\{{\cal D}_{3,3}\}$, violate the conditional CHSH inequalities $\langle{\rm CHSH}\rangle^{\rm con}_{c=1\,\rm opt}\leq 2$ and $\langle{\rm CHSH}\rangle^{\rm con}_{\rm opt}\leq 2$. We also illustrate that while states belonging to both  
$\{{\cal D}_{3,2}\}$, $\{{\cal D}_{3,3}\}$ can attain the maximum value $2\sqrt{2}$ for 
$\langle{\rm CHSH}\rangle^{\rm con}_{c= 1\,\rm opt}$, the maximum value for $\langle{\rm CHSH}\rangle^{\rm con}_{\rm opt}$ achievable by the states belonging to $\{{\cal D}_{3,2}\}$, $\{{\cal D}_{3,3}\}$ are $2.552$ and $2\sqrt{2}$ respectively. This observation lead us to define a quantity $\cal Q$ which quantifies nonlocality in pure symmetric states belonging to $\{{\cal D}_{3,2}\}$ and $\{{\cal D}_{3,3}\}$. Three-qubit GHZ state belonging to $\{{\cal D}_{3,3}\}$ is seen to have ${\cal Q}=1$ thereby exhibiting maximum nonlocality. On the same token, we have shown ${\cal Q}\approx 0.66$ for three-qubit W state belonging to $\{{\cal D}_{3,2}\}$. 
 Furthermore, we have shown that  symmetric three-qubit states belonging to the three distinct spinor class $\{{\cal D}_{3,3}\}$ maximally violate {\em six}  of the 46 classes of tight Bell inequalities in the (3,2,2) scenario. There are no  tight Bell inequalities in the (3,2,2) scenario that are maximally  violated by  the two distinct spinor class $\{{\cal D}_{3,2}\}$. The nonlocality tests discussed here offer useful signatures to  distinguish the quantum states belonging to  the   inequivalent SLOCC classes $\{{\cal D}_{3,2}\}$ and $\{{\cal D}_{3,3}\}$.   We believe that our work motivates experimental tests on the violations of nonlocality by pure three-qubit symmetric states in different physical platforms.  
	\section*{Acknowledgements} We thank Professor A. K. Rajagopal for going through the manuscript and for making insightful suggestions. KA acknowledges financial support from UGC-RGNF, India. ASH is supported by the Foundation for Polish Science 
	(IRAP Project, ICTQT, contract no. 2018/MAB/5,  co-financed by EU within Smart Growth Operational Programme). HSK acknowledges the support of NCN through grant SHENG (2018/30/Q/ST2/00625). This work was partly done when HSK was at The Institute of Mathematical Sciences, Chennai, India and progressed further at ICTQT, Gdansk, Poland. Sudha and ARU are supported by the Department of Science and Technology(DST), India through Project No. DST/ICPS/QUST/Theme-2/2019 (Proposal Ref. No. 107).

\end{document}